\documentclass[10pt]{article}
\usepackage{graphicx,floatflt,amssymb,epsfig,epsf} 
\def\be {\begin{equation}}
\def\ee {\end{equation}}
\def\bea {\begin{eqnarray}}
\def\eea {\end{eqnarray}}

\def\ztwo {{\cal Z}_2}

\def\pir {\pi R}
\def\stwo {\sqrt{2}}
\def\spir {\sqrt{\pir}}
\def\mhsbar {\overline{m_h^2}}

\def\opcit(#1){ {\em op. cit.}, #1}

\def\issue(#1,#2,#3){#1 (#3) #2} 

\def\APP(#1,#2,#3){Acta Phys.\ Polon.\ \issue(#1,#2,#3)}
\def\ARNPS(#1,#2,#3){Ann.\ Rev.\ Nucl.\ Part.\ Sci.\ \issue(#1,#2,#3)}
\def\CPC(#1,#2,#3){Comp.\ Phys.\ Comm.\ \issue(#1,#2,#3)}
\def\CIP(#1,#2,#3){Comput.\ Phys.\ \issue(#1,#2,#3)}
\def\EPJC(#1,#2,#3){Eur.\ Phys.\ J.\ C\ \issue(#1,#2,#3)}
\def\EPJD(#1,#2,#3){Eur.\ Phys.\ J. Direct\ C\ \issue(#1,#2,#3)}
\def\IEEETNS(#1,#2,#3){IEEE Trans.\ Nucl.\ Sci.\ \issue(#1,#2,#3)}
\def\IJMP(#1,#2,#3){Int.\ J.\ Mod.\ Phys. \issue(#1,#2,#3)}
\def\JHEP(#1,#2,#3){J.\ High Energy Physics \issue(#1,#2,#3)}
\def\JPG(#1,#2,#3){J.\ Phys.\ G \issue(#1,#2,#3)}
\def\MPL(#1,#2,#3){Mod.\ Phys.\ Lett.\ \issue(#1,#2,#3)}
\def\NP(#1,#2,#3){Nucl.\ Phys.\ \issue(#1,#2,#3)}
\def\NIM(#1,#2,#3){Nucl.\ Instrum.\ Meth.\ \issue(#1,#2,#3)}
\def\PL(#1,#2,#3){Phys.\ Lett.\ \issue(#1,#2,#3)}
\def\PRD(#1,#2,#3){Phys.\ Rev.\ D \issue(#1,#2,#3)}
\def\PRL(#1,#2,#3){Phys.\ Rev.\ Lett.\ \issue(#1,#2,#3)}
\def\SJNP(#1,#2,#3){Sov.\ J. Nucl.\ Phys.\ \issue(#1,#2,#3)}
\def\ZPC(#1,#2,#3){Zeit.\ Phys.\ C \issue(#1,#2,#3)}

\begin{document} 
\begin{flushright} 
CU-PHYSICS/07-2007\\
\end{flushright} 
\vskip 30pt 
 
\begin{center} 
{\Large \bf Production of Higgs boson excitations of universal extra
dimension at the Large Hadron Collider}\\
\vspace*{1cm} 
\renewcommand{\thefootnote}{\fnsymbol{footnote}} 
{\large {\sf Biplob Bhattacherjee} and {\sf Anirban Kundu} } \\ 
\vspace{10pt} 
{\small 
   {\em Department of Physics, University of Calcutta, 92 A.P.C. 
        Road, Kolkata 700009, India}}
 
\normalsize 
\end{center} 
 
\begin{abstract} 
The Kaluza-Klein excitations of the Higgs bosons of the universal extra
dimension model are extremely challenging to detect. We discuss the 
production and possible detection mechanisms of such excited scalars
at the LHC. The dominant production mechanism of such scalars is from
the decay of the excited third generation quarks. In particular, the charged
Higgs boson has a large production cross-section over most of the parameter
space. We highlight how one may detect these excited scalars. We also 
comment on the production and detection of excited neutral scalars.
 
\vskip 5pt \noindent 
\texttt{Keywords:~~Universal Extra Dimension, Large Hadron Collider, Charged
Higgs}
\end{abstract}

\renewcommand{\thesection}{\Roman{section}} 
\setcounter{footnote}{0} 
\renewcommand{\thefootnote}{\arabic{footnote}} 

\section{Introduction}

The Universal Extra Dimension (UED) scenario \cite{acd}, where all 
Standard Model (SM)
fields can propagate into the compactified extra dimension(s), has attracted
a lot of interest recently. In the minimal version of UED (hereafter called 
mUED) \cite{cms1}, there is only one extra dimension $y$ 
compactified on a circle
of radius $R$. Every SM particle is associated with an infinite but discrete
tower of similar particles, the $n$-th level (this will be called the 
Kaluza-Klein, or KK, number) of which has a tree-level mass of $\sqrt{m_0^2+
n^2/R^2}$, where $m_0$ is the mass of the SM particle. Since momentum along
the fifth dimension is conserved, so is the KK number. 

To get chiral fermions in 4-dimensional theories, one needs an $S_1/\ztwo$
orbifolding, identifying $y$ with $-y$ in the interval $-\pi R \leq y <\pi R$.
The KK number, however, can be violated radiatively. Such loop diagrams are
divergent and one needs to introduce suitable counterterms located at the
fixed points $y=0,\pi R$ to cancel those divergences. These counterterms
depend on the effective cut-off of the theory, $\Lambda$, which, for any
realistic model, should be much higher than $R^{-1}$. 

The KK particle masses and mixing matrices are modified because of finite
corrections coming from Lorentz invariance violating loops and log-divergent
($\sim\ln\Lambda^2$) contributions coming from terms located at the fixed 
points (the so-called boundary corrections) \cite{cms1,georgi,carena}.
This causes significant splitting among the particle masses of any KK level
and has important effects on collider phenomenology. It is also possible
to violate the KK number by two units, though the KK-violating terms are of
much weaker strength. The KK-parity, defined as $(-1)^n$, is still conserved,
and the lowest $n=1$ particle is stable. For all practical purposes, the
lightest KK particle (LKP), $B_1$, is the excitation of the 
hypercharge gauge boson
$B$, and is an excellent cold dark matter (CDM) candidate \cite{servant}. 
The data on CDM
density translates into an upper limit of $R^{-1}$, which is about 1 TeV
\cite{hooper}, so that the universe is not overclosed. We do not consider
the gravitons in the mUED model. 

The importance of collider phenomenology for TeV-scale 
extra dimensional models is well-known \cite{antoniadis}.
The signatures of mUED in $e^+e^-$, $p\bar{p}$, and $pp$ colliders 
have been extensively studied \cite{lhc,ilc,self}.
Data on low-energy observables indicate that $1/R > 250$-300 GeV 
\cite{acd,debrupa,buras,low-rest}. An analysis of the process $B \to
X_s+\gamma$ suggests that it can be even higher, about 600 GeV \cite{haisch}.
The question whether mUED can be discriminated from supersymmetric models
with a similar mass spectrum \cite{cms2} has apparently been resolved with
the answer being positive \cite{ilc,webber}. The signals for
excited quark and lepton production and their subsequent decay are quite
clean, in the sense that there are only two relevant parameters, $R^{-1}$
and $\Lambda$, and so the predictive power of the model is high. 
The so-called ``smoking gun" signal of mUED, namely, the $n=2$ gauge boson 
production, has also been investigated in the context of the Large Hadron
Collider (LHC) \cite{datta} and the International Linear Collider (ILC)
\cite{self,biplob-lcws}.

While the mUED model is completely parametrised by $R^{-1}$ and $\Lambda$, 
one may add a mass-like term $\mhsbar$ for the scalars, which is situated
symmetrically at the two fixed points $y=0$ and $y=\pi R$. Since it is the
excitation of the Higgs sector that we are interested about, we will keep
this term as a free parameter. So, strictly speaking, we are considering
the mUED model expanded to include the scalar mass term $\mhsbar$. 
The effect of this term was
never seriously investigated, apart from a study in the context of the ILC
\cite{self-higgs}. 
The term affects only the masses of the Higgs boson excitations, and
hence the production and decay of those scalars. 

Once the excited leptons, quarks, and gauge bosons are discovered at the
LHC, it becomes mandatory to explore the Higgs sector, not only to
complete the UED spectra but also to have an idea of $\mhsbar$, assuming
that the SM Higgs boson would be discovered and precisely studied by
then. Unfortunately, it seems a major challenge to the experimentalists
to detect these excited scalars \cite{self-higgs}. The reason is that
the signal, for most part of the parameter space, is one or more very
soft $\tau$-leptons, often below the detection limit of the ATLAS or
the CMS detectors. However, it is not impossible, and we discuss in this
letter, albeit qualitatively, why the task is challenging and how one
should address the question. 
It goes without saying that for this study to be meaningful,
LHC must first discover and identify mUED, through the detection of 
excited leptons, quarks, and gauge bosons, so that one has at least a rough
idea about $1/R$ and maybe about $\Lambda$. 

The paper has been arranged as follows. In Section II, we briefly review
the mUED Higgs sector and the third generation quark sector
at the $n=1$ level. We show the possible production
and decay modes of these Higgs bosons. 
In Section III, we obtain the cross-section of the charged scalar at the
LHC. The signal as well as possible backgrounds are identified. We highlight
the major challenge in detecting the signal, and the way to overcome the
challenge by a careful study of the polarisation of the final-state $\tau$ 
lepton.
In Section IV we take up the study for neutral scalars. We find that
this sector remains a challenge even at the LHC. We comment and conclude 
in Section V.

\section{Scalars and Fermions of mUED}
\subsection{The mUED Higgs Bosons}  

In mUED, a five-dimensional field can be Fourier expanded as
\bea
\phi_+(x^\mu,y) &=& \frac{1}{\spir} \phi_+^{(0)}(x^\mu) + \frac{\stwo}{\spir}
\sum_{n=1}^\infty \cos\frac{ny}{R} \phi_+^{(n)} (x^\mu),\nonumber\\
\phi_-(x^\mu,y) &=& \frac{\stwo}{\spir}
\sum_{n=1}^\infty \sin\frac{ny}{R} \phi_-^{(n)} (x^\mu),\eea
where $\phi_+$ is even and $\phi_-$ is odd under $\ztwo$. Fields which are
odd under the $\ztwo$ orbifold symmetry do not have zero modes. Only even
fields have zero modes, which are identified with the SM particles.
The scalar fields are $\ztwo$-even, so are the first four components of the
gauge fields. 

The $n$-th level Higgs field is parametrised as
\be
H_n=\pmatrix{ \chi^+_n\cr \frac{h_n - i\chi^0_n}{\stwo} }
\ee
where $h_n$, $\chi^0_n$, and $\chi^+_n$ are the excitations of
CP-even neutral, CP-odd neutral, and charged scalars respectively (the
subscript refers to the KK number). There are three more colour neutral
scalars, which are the fifth components of the excitations of the
weak gauge bosons. These fields are $\ztwo$-odd and can occur first at the
$n=1$ level. Each of them mixes with the corresponding Goldstone excitations, 
and produce one Goldstone
at the excited level (which gets eaten up by the corresponding gauge boson
to make it massive). The other component remains in the physical
spectrum.

The Goldstone combinations are given by
\be
G^0_n=\frac{1}{m_{Z_n}}\left[ m_Z \chi^0_n-\frac{n}{R}Z^5_n\right],
\ee
and
\be
G^\pm_n=\frac{1}{m_{W_n}}\left[ m_W \chi^\pm_n-\frac{n}{R}W^{5\pm}_n\right].
\ee
The orthogonal combinations are the physical scalar fields, and we will call
them $A^0_n$ and $H^\pm_n$ respectively. The excitation of the SM Higgs boson
will be denoted by $h_n$.

It is clear that if $1/R \gg m_{W,Z}$, the $n\not= 0$
Goldstones are essentially the
fifth components of the gauge bosons, whereas the physical scalars are the
excitations of the $n=0$ Goldstones and the $n=0$ Higgs boson. We will 
work in this limit only. It will be shown that only in the large $1/R$ limit 
one may expect to observe some signal events.

In the absence of radiative corrections, the tree-level masses of the excited
scalars are given by
\be
m_{h_n,A_n^0,H_n^\pm}^2= \frac{n^2}{R^2}+m_{h,Z,W^\pm}^2,
   \label{higgstree}
\ee
but this relation
is modified by radiative corrections, whose effect is simply to add
a universal term $\delta m_H^2$ to the right-hand side of eq.\ 
(\ref{higgstree}) \cite{cms1}. The radiative correction is given by
\be
\delta m_H^2 = \frac{n^2}{R^2}
\left[\frac{3}{2}g^2+\frac{3}{4}{g'}^2 -\lambda\right]
\frac{1}{16\pi^2}\ln\frac{\Lambda^2}{\mu^2} + \mhsbar,
   \label{radcorhiggs}
\ee
where $g'$ and $g$ are the $U(1)_Y$ and $SU(2)_L$ gauge couplings
respectively, and $\lambda$ is the self-coupling of the Higgs boson,
given by $m_h^2=\lambda v^2$ where $v=246$ GeV. $\Lambda$ is the
effective cutoff scale and $\mu$ is the regularisation scale. For $n=1$,
we put $\mu=1/R$.  The term
$\mhsbar$ is arbitrary; this is the boundary mass term for the excited
scalars, and is not {\em a priori} calculable.
Along with $1/R$ and $\Lambda$, $\mhsbar$ forms
the complete set of input parameters to specify our version of the minimal
UED model (of course, one needs to know the SM Higgs boson mass, $m_h$).

Note that in the presence of these fixed point located terms, the orbifold
$\ztwo$ is no longer a good quantum number. If the terms are symmetrically
located at the fixed points, only the KK-parity, defined as $(-1)^n$, is
conserved. Only the $\ztwo$-even states mix with each other, since the
wavefunctions for the $\ztwo$-odd states vanish identically at the fixed
points. Of course, all scalars are $\ztwo$-even, so the mixing is theoretically
relevant for us. If we wish to keep the orbifold $\ztwo$ as an almost good
quantum number, the mixing between different KK-levels should be small. 
For example, if we take $m_h=120$ GeV and $R^{-1}=500$ GeV, $\mhsbar=10^4$
GeV$^2$ gives a one percent mixing between $n=0$ and $n=2$ states. Let us
take this to be the limit and keep the magnitude of $\mhsbar$ to be less than
$10^4$ GeV$^2$.

Let us concentrate on the $n=1$ level and note a few points here 
following eqs.\ (\ref{higgstree}) and (\ref{radcorhiggs}).
\begin{itemize}
\item
The hierarchy $m_{h_n} > m_{A^0_n} > m_{H^\pm_n}$ is fixed. However, the
splitting among these levels is not.
\item
If we keep $R^{-1}$ and $\Lambda$ fixed, the scalar masses depend on 
$\mhsbar$ and $\lambda$. Thus, for larger SM Higgs mass ({\em i.e.}, for
larger $\lambda$), $H_1^\pm$ and $A_1^0$ masses go down if we keep
$\mhsbar$ fixed.
\item
On the other hand, $h_1$ will become more massive, because of the positive
$m_h^2$ contribution in eq.\ (\ref{higgstree}).
\item
For a fixed SM Higgs mass, all excited scalar masses increase with increasing
$\mhsbar$. Since $\mhsbar$ is arbitrary, it can even be negative. However,
for {\em large} negative values of $\mhsbar$, $H_1^\pm$ will become the
LKP, which is forbidden from astrophysical considerations. Thus, we have a
lower limit on $\mhsbar$, which is a function of $R^{-1}$, $\Lambda$, and
the SM Higgs mass $m_h$.
\end{itemize}

\begin{figure}[htbp]
\vspace{-10pt}
\centerline{
\rotatebox{-90}{\epsfxsize=9cm\epsfbox{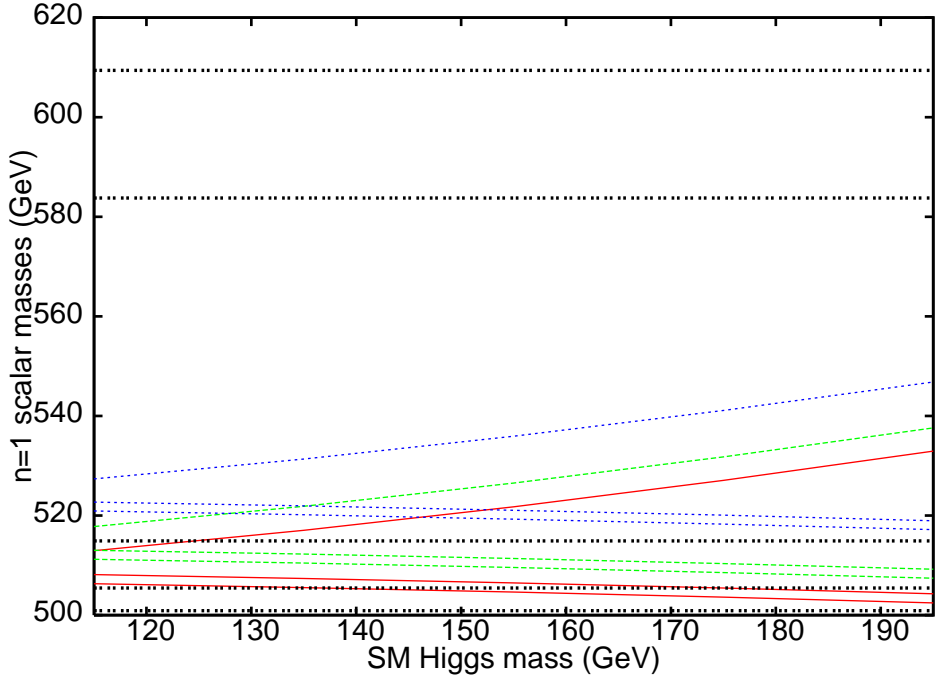}}}
\caption{The variation of $n=1$ scalar masses with the SM Higgs boson mass
$m_h$. The red, green and blue lines are for $\mhsbar=-5000$, 0, and
10000 respectively. In any bunch, the lines from bottom to top are
for $H_1^\pm, A_1^0$ and $h_1$. We have set $1/R=500$ GeV 
and $\Lambda R = 20$. The thick double-dotted horizontal lines denote,
from bottom to top, the masses for the LKP, $\tau^{(1)}$ (dominantly singlet),
$\tau^{(2)}$ (dominantly doublet), $t^{(1)}$ and $t^{(2)}$ respectively.}
    \label{higgsmass}
\end{figure}
The variation of the scalar masses is shown in fig.\ \ref{higgsmass}.
Since the KK-parity is conserved, the scalars
must decay leptonically, as all $n=1$ quarks are heavier than these scalars. 

\begin{figure}[htbp]
\vspace{-10pt}
\centerline{
\rotatebox{-90}{\epsfxsize=9cm\epsfbox{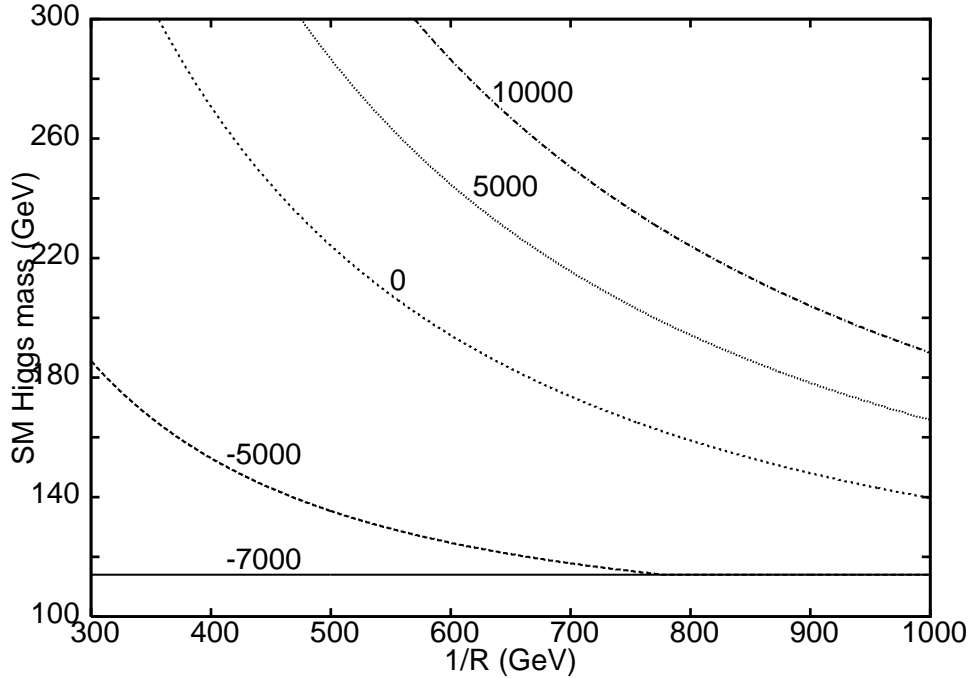}}}
\caption{The parameter space for two-body and three-body decays of $H^\pm_1$.
The lines are drawn for different values of $\mhsbar$, as shown in the plot.
Only three-body (and higher) decays are allowed above each line, while
below them, two-body decays are the dominant ones.
The lower horizontal line for $\mhsbar=-7000$ GeV$^2$ is for the
experimental limit of the Higgs boson, at 114 GeV. For more details, see text.}
   \label{twobody}
\end{figure}

\subsection{The mUED Fermions}

The $n=1$ fermions can be both $\ztwo$-even (left doublet and right singlet)
and $\ztwo$-odd (left singlet and right doublet). These states are mass
eigenstates for all fermions, except for the third generation quarks.
In the doublet-singlet basis, the top quark mass matrix is written as
\be
\pmatrix{1/R+\delta m_2 & {m_t}\cr
m_t & -1/R-\delta m_1}
\ee
where $\delta m_1(\delta m_2)$ are the radiative corrections for the singlet
(doublet) fields. Their expressions can be found in \cite{cms1}.
After diagonalisation and a chiral rotation, one gets the mass eigenstates
$t^{(1)}$ and $t^{(2)}$, where $t^{(1)}$ ($t^{(2)}$) is dominantly singlet 
(doublet), the composition being somewhere
between 97-100\%, depending upon the value of $1/R$. $t^{(2)}$
is slightly more massive than $t^{(1)}$ and hence we expect a marginally larger
cross-section for $t^{(1)}$ pair production than that for $t^{(2)}$ at the LHC.

The same mechanism works for the leptons. We will be interested in $n=1$
$\tau$ leptons, for which the off-diagonal terms are much smaller than those
for the top quark. Still, this makes $\tau^{(1)}$, the dominantly singlet
one, the lightest charged $n=1$ lepton. We refer the reader to fig.\
\ref{higgsmass} for an idea about the mass spectra of the LKP, $\tau^{(1),(2)}$,
and $t^{(1),(2)}$.

\subsection{Decay of $n=1$ Scalars}

For most of the parameter space $H^+_1$ decays
into $\tau^+_1\nu_{\tau 0}$ and $\tau^+_0\nu_{\tau 1}$ (the subscripts stand
for the KK levels). The $\tau_1$ will cascade down to $\tau_0$ plus LKP
(similarly for $\nu_{\tau 1}$), so the final state is a soft $\tau$ plus
large missing energy. However, this channel may not always be kinematically 
open. With increasing $R^{-1}$ (keeping $m_h$ fixed), or with
increasing $m_h$ (keeping $R^{-1}$ fixed), $H^+_1$ comes closer to $\tau_1$
and ultimately goes below it. This closes the two-body channel,
leading to three-body channels
$H^+_1\to f \bar{f'}$ plus LKP, where $f$ and $f'$ are $n=0$ fermions. 
Even before this takes place, the final state $\tau$ becomes so soft as
to miss detection and $H^+_1$ decays invisibly. 
The point where the transition takes place is shown in fig.\ \ref{twobody}
as a function of $\mhsbar$. For example, if $R^{-1}=1$ TeV and $\mhsbar=0$,
the two-body channel will be open only if $m_h < 140$ GeV. This can also
be guessed from fig.\ \ref{higgsmass} by looking at the mass difference
of $H_1^\pm$ and $\tau^{(1)}$, which should be more than $m_{\tau_0}$ for
two-body channels to remain open \footnote{There is a very narrow region
in the parameter space where the $\tau$ channels close but the channel 
$H_1^+\to \mu^{(1)} \nu_{\mu 0}$ remains open. We do not spend any further
time on that region, since such soft muons will definitely be missed by
the detector. However, fig.\ \ref{twobody} is drawn taking the muon channel
into account. }. 

We consider only the two-body decays of $H^+_1$.
Note that the charged Higgs is still short-lived enough to decay
within the detector. 

For all practical purposes, $n=1$ neutral Higgs bosons $h_1$ and $A^0_1$
decay into one $n=1$ and one $n=0$ $\tau$ lepton (charged or neutral), 
as long as the tree-level two-body
channel remains open. When this becomes kinematically forbidden, they decay
into a photon and the LKP $B_1$, the amplitude being dominated by the top quark
loop. Again, we consider only the $\tau$ signal in the final state. 

It should now be clear why the detection of such Higgs bosons is going to be
a major challenge. All of them ultimately decay to one or two $\tau$ leptons,
but they are going to be soft since most of the energy is carried away by the
LKP. The detection limit $p_T^\tau > 20$ GeV, which we take to be the case
for CMS and ATLAS, removes almost all the signal events. Thus the scalars may go
undetected, even though the production cross-section is large. 
This is in sharp contrast to the charged Higgs signals in the $\tau$ channel
in supersymmetry or two-Higgs doublet models, where the daughter $\tau$
must be hard and easily detectable. 
Note that when the two-body channels are closed, the $n=0$
fermion-antifermion pair must be softer, since the mass splitting between
the scalars and the LKP goes down. 

\section{$H^\pm_1$ at the LHC}
\subsection{Production}

Let us first study the charged Higgs boson production at the LHC. The
projected annual luminosity is about 100 fb$^{-1}$; throughout the study
we take this as the benchmark luminosity.
We will vary $R^{-1}$ and go upto $R^{-1}=1$ TeV, mostly because
the dark matter density reaches the overclosure bound, and also for the
simple fact that beyond this, the number of final soft-$\tau$ events after
all the kinematic cuts is too low for detection. 

The dominant production mechanism is through the real production of $n=1$
top pair $pp\to t^{(1)}{\bar{t}}{}^{(1)}$. 
All scalars are $\ztwo$-even, and so are all
$n=0$ particles, so only the decay of the $\ztwo$-even $t^{(1)}$ matters. 
This is dominantly SU(2) singlet, and so will decay almost entirely to
$b_0H_1^+$ (the $\ztwo$-odd $t^{(1)}$ will also decay through the same channel
after a vectorial mass insertion). At the leading order (LO), the gluon-gluon
fusion is the dominant mechanism over $q\bar{q}$ fusions.
However, the next-to-leading order (NLO) contributions should be significant.
Even the pure QCD contributions depend on $1/R$ \footnote{
One notes that the ${\cal O}(\alpha_s^3)$ NLO corrections are not exactly
identical to those for $t\bar{t}$ production. For example, in the one-loop
virtual correction to the $gg\to t^{(1)}{\bar{t}}{}^{(1)}$ process (which,
by interference with the LO term, produces an ${\cal O}(\alpha_s^3)$
correction), the heavy $n=1$ gluons, apart from $n=0$ gluons, also participate.
In fact, to evaluate such terms, one needs to sum a series of contributions
coming from higher KK modes.}.
The NLO correction to the LO process is
yet to be computed; so we take $K$, defined as $K=\sigma(pp\to
t^{(1)}{\bar{t}}{}^{(1)})_{NLO}/\sigma(pp\to t^{(1)}{\bar{t}}{}^{(1)})_{LO}$, to
be the same as of the SM process $pp\to t\bar{t}$. This procedure, arguably,
is open to criticism; in particular, one may point out that $K$ should
be a function of $R^{-1}$, and as one goes to large $R^{-1}$, the relative
importance of $q\bar{q}$ fusion increases. 

\begin{figure}[htbp]
\vspace{-10pt}
\centerline{
\rotatebox{-90}{\epsfxsize=9cm\epsfbox{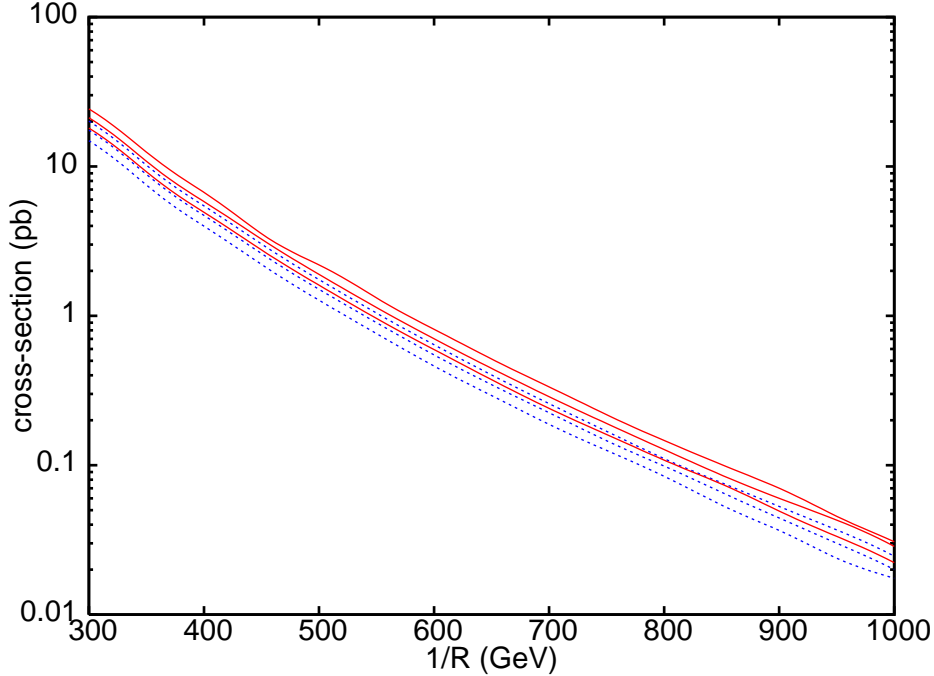}}}
\caption{$pp\to t^{(1)}\bar{t^{(1)}}$ (red solid lines) and 
$pp\to t^{(2)}\bar{t^{(2)}}$ (blue dotted
lines) cross-sections at the LHC. We have used $\Lambda R=20$, and the MRST
parton distribution at the NLO. The lines in each band, from top to bottom, 
are for QCD factorisation scale $\mu=0.5 R^{-1}$, $R^{-1}$ and $2R^{-1}$
respectively.}
   \label{topprodfig}
\end{figure}

We equate the regularisation and the factorisation scales $\mu_R=\mu_F=\mu$
(not to be confused with the $\mu$ in eq.\ (\ref{radcorhiggs})),
and evaluate the cross-section at three different points: $\mu=0.5R^{-1}$,
$R^{-1}$, and $2R^{-1}$. This, hopefully, makes the result more stable with
respect to the higher order corrections.  
CalcHEP v2.4.5 \cite{calchep} is used for calculation of the cross-sections
as well as event distributions. The minimal UED model has been
added to the basic CalcHEP kernel by us. (The necessary files may be obtained
by writing to one of us; however, PYTHIA interfacing is not yet implemented.)
We use the MRST parton distribution at the next-to-leading order (NLO) level.
The $K$-factors, for the three choices of $\mu$ mentioned above, are
1.23, 1.43, and 1.59 respectively \cite{bcmn}. 
We expect a further 20\% uncertainty from different choices of the 
parton distribution functions \cite{wagner}. The 
production cross-sections for $t^{(1)}$ and $t^{(2)}$ pairs are plotted
in fig.\ \ref{topprodfig}.

All subdominant $H_1^\pm$ production
mechanisms (even when taken together), like vector boson fusion or $gg\to
H_1^+H_1^-$ through quark box, have smaller 
cross-sections than the inherent QCD uncertainty of real $n=1$ top pair
production; one must remember that two $n=1$
particles must be produced together. For this reason we concentrate {\em only}
on the production of $H_1^\pm$ through real $t^{(1)}$ decay. 
In fig.\ \ref{topprodfig}, 
we show the production cross-section of both $t^{(1)}_1$ and $t^{(2)}_1$
pairs as a function of $R^{-1}$ with $\Lambda R=20$; at this point the value
of $\mhsbar$ is irrelevant. The production is through strong interaction but 
the $t^{(2)}$ pair production cross-section is slightly smaller due to its
higher mass. The lighter top $t^{(1)}$ decays almost entirely to $b_0H^+_1$ 
and hence to the $\tau\nu_\tau$ channel.


\subsection{Decay and Background}

If we stick to the two-body decay channels of $H_1^\pm$,
the decay mode $H_1^+\to W_1^+ + \gamma$ opens only at very high $\mhsbar$
($\sim 10^4$ GeV$^2$) and quickly becomes the only significant channel, but
before that $H_1^+ \to \tau_1^+\nu_{\tau 0}$ is the dominant channel and
$H_1^+\to \tau_0^+ \nu_{\tau 1}$ is the subdominant one. The latter has a
branching fraction of about 10\% for $R^{-1}=300$ GeV, but quickly drops to
zero for larger $R^{-1}$. 
For this work, we will
assume $t^{(1)}$ to decay entirely through the $H_1^+ \to \tau_1^+\nu_{\tau 0}$
channel, the $\tau_1$ cascading down to a soft $n=0$ $\tau$ lepton and missing
energy carried away by the LKP. Thus, the $t^{(1)}{\bar{t}}^{(1)}$ pair 
decays down
to a pair of oppositely charged soft $\tau$s, a pair of hard $b$ jets and 
a large missing energy. Since these soft $\tau$s are the true signature of
the charged Higgs, one must detect at least one of them. 
{\em This gives our signal: two hard $b$-jets, at least
one $\tau$ with $p_T<30$ GeV, and a large missing energy.} We select only
those events which have missing $p_T$ between 150 and 300 GeV. While the
lower cut removes the SM background almost completely (only 417 events 
survive, this obviously does not depend upon $1/R$),
the upper cut reduces the background 
coming from other UED processes.
For detection, $p_T$ of the $\tau$(s) should be more than 20 GeV.

There are two possible sources of large background for this event.
Both come from UED processes.  The first one comes from 
$pp\to t^{(2)} {\bar{t}}{}^2$, $t^{(2)}\to b_0W_1$ 
($b_1 W_0$ channel is kinematically forbidden), $W_1\to \tau_0\nu_{\tau_1}$
or $W_1 \to \tau_1\nu_{\tau_0}$. For our future reference, we call the
process where both $\tau$s come from $W_1 \to \tau_1$ ($W_1\to\tau_0$)
as BG1a (BG1b), and the process where one $\tau$ comes from $W_1\to\tau_1$
and the other from $W_1\to\tau_0$ as BG1c. 
The second one comes from the process $pp \to b^{(2)}\bar{b^{(2)}}$, 
where $b^{(2)}$ is the $n=1$ $b$ quark which is dominantly $SU(2)$ doublet. 
Note that $b^{(2)}$ decays mostly to $b_0 Z_1$ (85-91\%),
and $b_0 B_1$ (9-15\%), and only about 0.1\% of the times
to $b_0 h_1/A^0_1$.  The variation
is due to the shifting of levels for different values of $1/R$. 
There are three subprocesses for this. 
The first subprocess is where one $b^{(2)}$
decays to $Z_1$ which in turn decays to $\tau_0\tau_1$, and the other
$b^{(2)}$ decays to $B_1$.  We call this BG2a.
This gives the $2b+2\tau+p_T\!\!\!\!\!\!/$ ~background.
In the second subprocess (BG2b), both $b^{(2)}$ go to $b_0 Z_1$, 
and while one $Z_1$ decays to $\tau_0\tau_1$,
the other decays to $\nu_0\nu_1$. The $\nu_1$ subsequently decays to
$\nu_0$ and $B_1$, so this is invisible. The third subprocess, BG3c, is
the one where both $b^{(2)}$ go to $b_0 Z_1$, ultimately giving rise to
a $2b+4\tau+p_T\!\!\!\!\!\!/$ ~background, but 2 or 3 $\tau$s are missed
as their $p_T$ fall below the 20 GeV cut, and none of them are harder
than 30 GeV. 
The $s$-channel $\tau_1$ pair production cross-section, along with two 
$b$-jets, is much smaller and may be neglected. 

Both $W_1$ and $Z_1$ decay only through leptonic channels. About one-third
of the times $W_1$ decays to $\tau\nu_\tau$. $Z_1$ has an approximately $50\%$ 
branching ratio to the neutrinos, and about $1/6$-th of the time it goes to
$\tau_0\tau_1$. Quark channels are kinematically forbidden.
It is easy to make a rough estimate of
signal and background events without any cut. Let $N_d(N_s)$ be the number of
events for pair production of any $n=1$ third generation quark which is
dominantly $SU(2)$ doublet (singlet). Since the splitting is small for top and
almost zero for bottom, $N_d\approx N_s = N$ (the production is through QCD
process, which is chirality-blind). The number of signal events is $N$,
whereas the number of $W_1$ background events is $N/9$. The number of
$Z_1$ background events (first subprocess) is roughly 
$\frac{2}{6}Nf_Z(1-f_Z)$ where $f_Z$ is
the branching fraction of $b^{(2)}$ to $b_0 Z_1$ (the factor of 2 is due to 
combinatorics). Putting the numbers, this comes out to be $0.070$-$0.085 N$.  
For the second subprocess, it is roughly $N f_Z^2/6$.
We assume that electrons or muons coming from $Z_1$ or $W_1$ decay have 100\%
detection efficiency.

\begin{table}[htbp]
\begin{center}
\begin{tabular}{||c|c|c|c|c|c|c|c|c||}
\hline
$R^{-1}$ & Signal & BG1a  & BG1b & BG1c & BG2a & BG2b & BG2c & Total\\
(GeV)    & events & events&events&events&events&events&events&Background\\
\hline
500      & 42     & 77 & 170 & 188 &  92 &  265 & 136 &  928  \\
600      & 67     & 68 & 93  & 161 &  26 &  351 &  90 &  789  \\
800      & 60     & 24 & 23  &  46 &  12 &  178 &  19 &  302   \\
1000     & 54     &  6 &  6  &  12 &   4 &   56 &   5 &   89   \\
\hline
\end{tabular}
\caption{Number of signal and background (dominant) events, with at least
one $\tau$ in the final state, coming from gluon-gluon fusion. 
For the definition of the background processes, and the cuts applied, see
text. One must add, with the last column,
 417 background events coming from SM top pair
production. The background
can be further reduced from $\tau$-polarisation studies (see text).}
\end{center}
\end{table}

In Table 1 we show the number of events for the signal process of $H_1^+$
production and for the dominant background processes. The events were
generated by CalcHEP and we assume a $b$ detection efficiency of 100\%.
Only the $p_T$ cut (20 GeV $< p_T<$ 30 GeV) was applied on each $\tau$,
and the $b$ jets are assumed to have a $p_T>20$ GeV.
The $\tau$ events can only be observed if $1/R$ is large --- for small
$1/R$, the separation between $\tau_1$ and the LKP is so small (this
does not depend on $\mhsbar$ but mildly depends on $\Lambda$, we have taken
$\Lambda R = 20$) that the resultant $n=0$ $\tau$ is too soft to be detected.
Even with large $1/R$, the background is larger than the signal, though
the significance is not negligible. However, one should be able to sharpen
the signal by using the polarisation of the one-prong decays of the $\tau$
\cite{dp99}.

If we consider one-prong hadronic decays $\tau^-\to (\pi^-,\rho^-,a_1^-) 
\nu_\tau$, the distribution of the final-state mesons depend upon whether
the $\tau$ came from $H_1^+$ (hence dominantly right-chiral) or
from $W_1^+$ (hence dominantly left-chiral). For $\rho$ and $a_1$, their
longitudinal and transverse modes can be separated by looking at the
difference of $p_T$ between the two pions that come out of their decay. 
We estimate, from existing studies, that the significance of the signal
may be around the $3\sigma$ confidence level by using the $\tau$-polarisation
method. The second reference in \cite{dp99} does a thorough job for hard
$\tau$s, by event generation and full detector simulation, and finds
that the relevant cross-section (cross-section times the efficiencies
for one-prong $\tau$ detection and $b$-tagging) for signal and background
(coming from $t\bar{t}$ pair production) processes are 1.1 fb and 0.15 fb,
respectively. Stricyly speaking, this is not applicable here, because an
important ingredient of their study is the hardness of the $\tau$ jet:
they have chosen $E_T^\tau > 100$ GeV, whereas we must confine ourselves
to low-energy $\tau$s. 
However, we also stress that this paper is more of a qualitative
nature, and a detailed quantitative study, with full detector simulation,
should be taken up. 

\section{Production of $h_1$, $A^0_1$ at the LHC}

\begin{figure}[htbp]
\vspace{-10pt}
\centerline{
\rotatebox{-90}{\epsfxsize=6.0cm\epsfbox{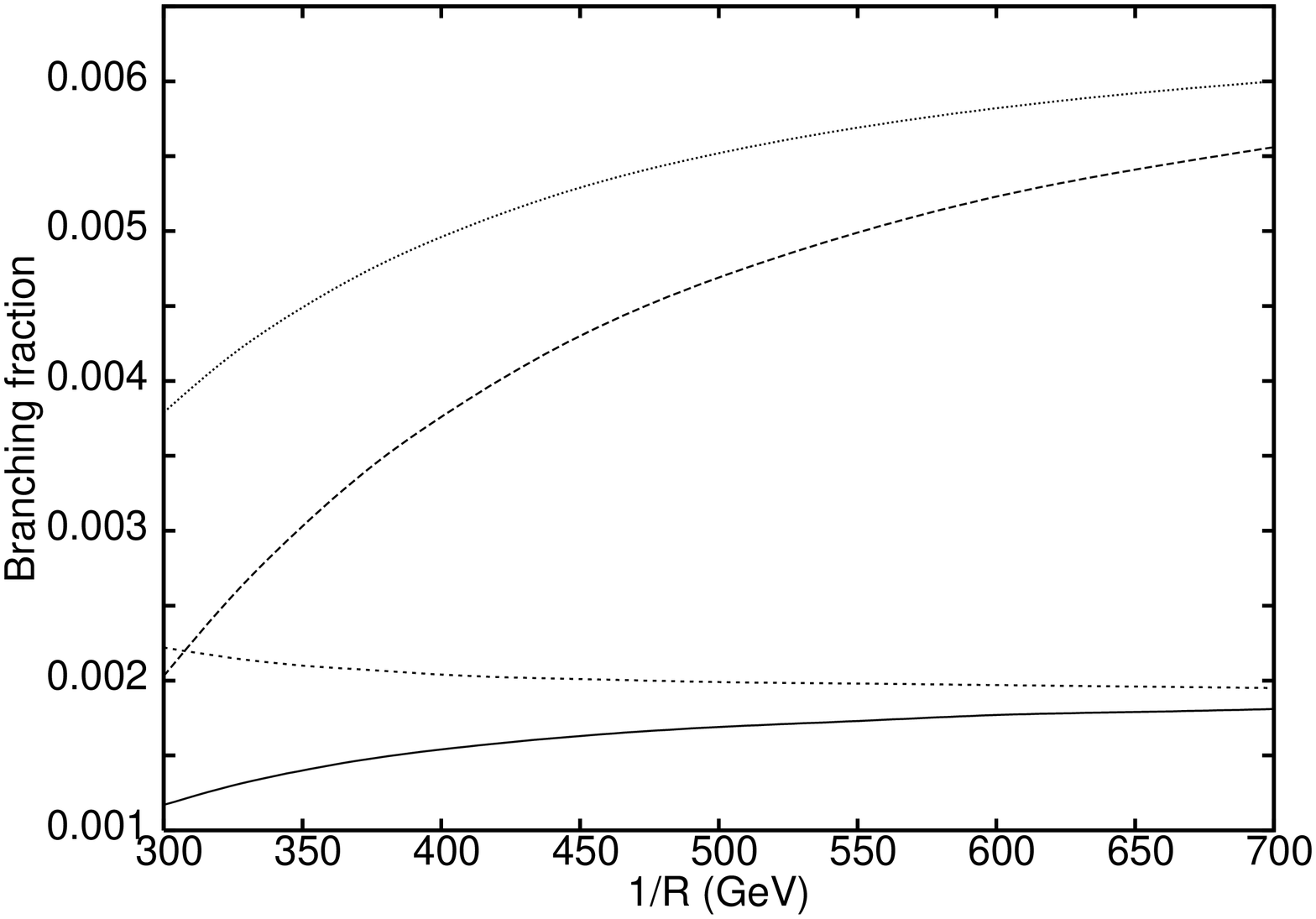}}
\hspace{-2mm}
\rotatebox{-90}{\epsfxsize=6.0cm\epsfbox{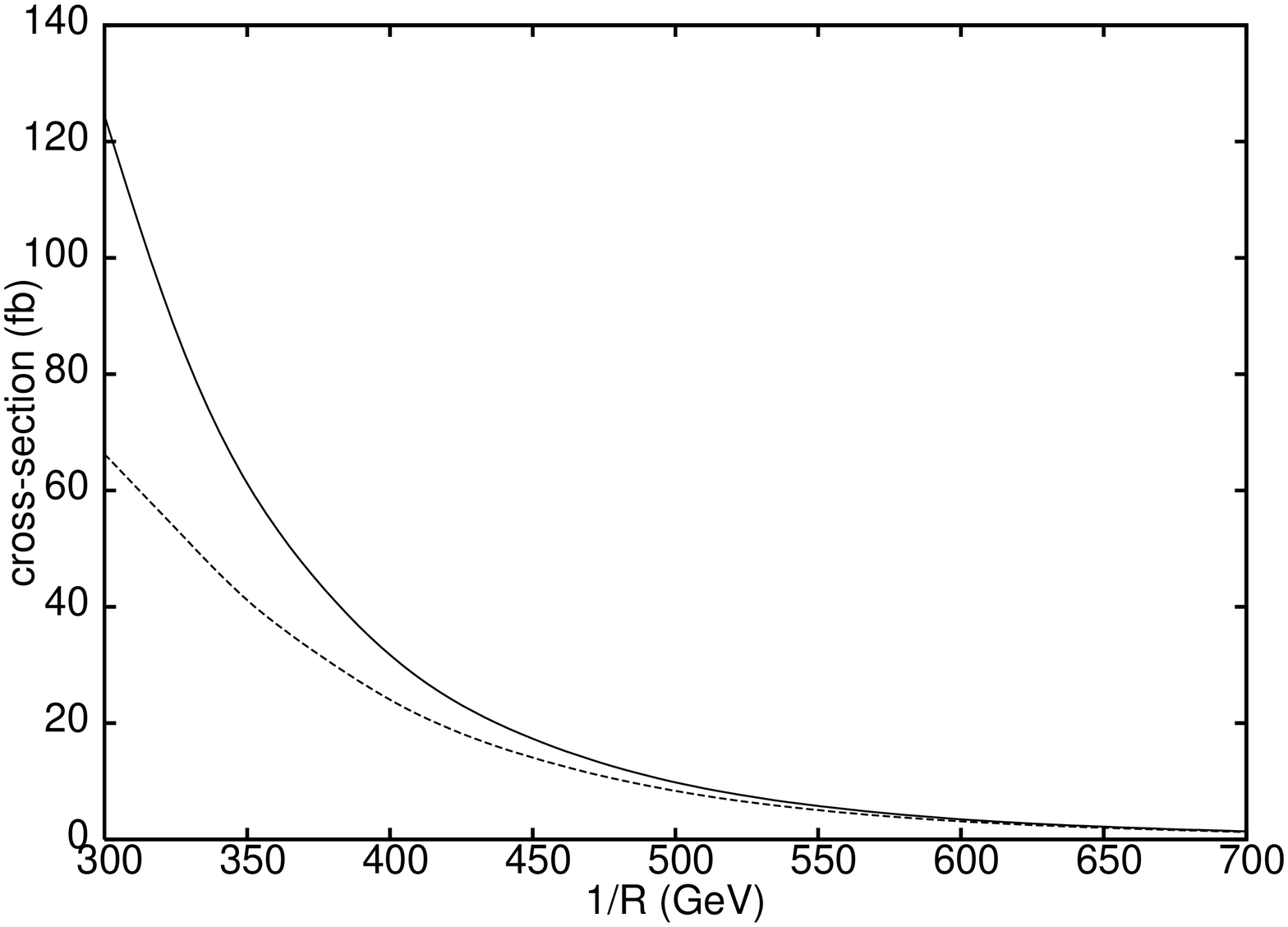}}}
\caption{(a) Branching fraction for neutral Higgses. From top to bottom,
at the right-hand edge, are: $b^{(1)}\to A^0_1$, $b^{(1)}\to h_1$,
$b^{(2)}\to A^0_1$, $b^{(2)}\to h_1$. Note that contrary to the top system,
$b^{(1)}$ is heavier. (b) Production cross-section for neutral Higgses. The
upper (lower) curve is for $A^0_1$ ($h_1$).}
    \label{neutprod}
\end{figure}

The mUED model contains two $n=1$ neutral Higgs bosons: $h_1$ and $A^0_1$.
There are three major processes for their production:
(i) Bjorken process $V_0^\ast\to V_1 h_1(A^0_1)$, where $V$ is a generic
gauge boson; (ii) electroweak vector-boson fusion; and 
(iii) associated production $pp\to b_1{\bar{b}}_1$, $b_1\to b_0 h_1(A^0_1)$. 
It can be even intuitively understood, and numerically confirmed, that
the last process will be the dominant one. The Bjorken process suffers from
excessive off-shellness of $V_0$, which is basically an $s$-channel
suppression. The vector-boson fusion is also suppressed, since the initial
bosons must be $n=0$ (radiation of $n=1$ bosons have negligible chance)
and one must produce two $n=1$ states in tandem. The third process, though
the dominant one, suffers from a roughly $m_b^2/m_t^2$ suppression 
over the $H_1^+$ production cross-section. In fig.\ \ref{neutprod}(a) 
we show the branching
fractions of $b^{(1)}$ and $b^{(2)}$ to $h_1$ and $A^0_1$; all of them
are small. The production cross-section for $h_1$ and $A^0_1$ is shown in 
fig.\ \ref{neutprod}(b). The difference is mainly due to the kinematic factors.  

The neutral Higgses decay, wherever kinematically possible, almost
entirely to $\tau_0\tau_1$. For $b^{(1)}$ pair production, the signal
will be where one $b^{(1)}$ decays to a neutral Higgs and the other decays
to $b_0 B_1$. If we stress on the detection of at least one $\tau$, which
is the best that we can do, the signal will be two hard $b$ jets plus 
one $\tau$ plus large missing energy, which is identical to the signal
of $H_1^+$. Therefore the backgrounds will be identical as discussed
earlier. However, the signal cross-section is smaller than the charged
Higgs cross-section by at least two orders of magnitude, so even with
the use of $\tau$ polarisation, detection of any excess event over the
charged Higgs signal has a pretty bleak prospect.

One can also produce $b^{(2)}$ pairs. Here one $b^{(2)}$ decays to
$Z_1$; thus, a possible signal event can be $2b+2\tau+2\mu$ plus missing
energy, where all the leptons are soft. While this signal can hardly 
come from charged Higgs production, a significant background is the
production of two $Z_1$s, one going to $\tau^+\tau^-$ and the other to
$\mu^+\mu^-$, accompanied by missing energy. Even here, after the $p_T$
cut is applied, the signal becomes miniscule to the background, but
one may try to apply the $\tau$ polarisation technique to extract the
neutral Higgs signal. 

If the neutral Higgs (in particular $A^0_1$, which is always slightly
lighter than $h_1$) cannot decay to $\tau_0\tau_1$, it will have a 
two-body loop decay, going into a photon and the LKP. The signal will be
entirely different --- two hard $b$ jets plus a very soft photon plus
large missing energy. There are a number of possible backgrounds that
may swamp the signal, most dominant being the $b^{(1)}$ pair production
with an initial or a final-state soft photon.

\section{Conclusions} 

At the first excited level, the minimal UED model contains three scalars:
one charged, and two neutral. Their masses depend on four parameters:
$1/R$, $\Lambda$, $\mhsbar$, and $m_h$, the SM Higgs mass. Just like any
other KK excitation, the scalar masses increase with increasing $1/R$ and
$\Lambda$, and also with increasing $\mhsbar$, a parameter on
which no other KK excitations depend. The mass of $h_1$ increases but the
masses of $H_1^\pm$ and $A_1^0$ decrease with increasing $m_h$.

We have discussed how one may try to detect
these scalars at the LHC. These Higgses can decay only leptonically ({\em i.e.},
only to $\tau$), and the spectrum dictates that the $\tau$s must be 
soft. This poses a serious challenge in their detection, since LHC should
not be able to detect $\tau$s with $p_T<20$ GeV. In fact, if the limit is
set higher, {\em e.g.}, $p_T^\tau > 40$ GeV, as done in a recent ATLAS study
\cite{atlas}, all signal events are certainly going to be missed or completely
swamped by the background. 

The charged Higgs $H^+_1$ can be copiously produced as the first-level
decay product of $n=1$ top quark which is dominantly singlet. The signal
will be two hard $b$ jets, at least one $\tau$ with $p_T<30$ GeV and
large missing energy. The detector limitation for identifying the $\tau$
removes the majority of the signal
events. For example, we do not envisage the observation of any signal below 
$1/R=450$ GeV (that is why we safely assumed the Higgses to be pure Goldstone
excitations). The major backgrounds come from the decay of $W_1$ and $Z_1$. 
However, $\tau$s coming from them are mostly left-chiral, while the
$\tau$s coming from the decay of $H^+_1$ are mostly right-chiral. 
One may significantly reduce the background and sharpen the signal by
looking at the distribution of final-state mesons in one-prong hadronic
decays of the $\tau$. 

The detection of neutral Higgses is far more challenging, simply because
their production rate is much smaller than that of the charged Higgs.
One possible way out may be to look for the signal $2b+2\mu+1(2)\tau$ plus
missing energy, where the $b$ jets are hard but all leptons are soft.
The background is still severe, but this is one place where one may
get encouraging results by applying the $\tau$ polarisation method.

We have assumed 100\% tagging efficiency for the $b$ jets. Realistically,
this should be something like 60\%. Both signal
and background event rates should be scaled down by $\epsilon^2$ where
$\epsilon$ is the $b$-tagging efficiency (so a factor of 0.36 for 60\%
efficiency). We have also assumed 100\%
efficiency for $e$ or $\mu$ detection. A more thorough study is needed
which would include a full detector simulation (here we have only
simulated the events upto the production of $\tau_0$) as well as the
simulation of the final-state decay products of $\tau$. 

In summary, we urge our experimental colleagues to look for these unusual
Higgs decay signals --- Higgses that appear hadrophobic in nature and
to complicate the matter, give only soft $\tau$s. If we are stuck in some
unfavourable corner of the parameter space, all Higgses may be invisible.
Even if they are not, this is probably one of the stiffest challenge to the
Higgs hunters.

\centerline{\bf {Acknowledgements}}

We acknowledge helpful discussions with Amitava Datta, Anindya Datta,
Aseshkrishna Datta, Alexander Pukhov, and D.P. Roy. AK was supported by 
the Project SR/S2/HEP-15/2003 of DST, Govt.\ of India. BB was supported
by a research fellowship of UGC, Govt.\ of India.

\end{document}